\documentclass[onecolumn,preprint,aps,prl, floatfix,amsmath,amssymb,superscriptaddress]{revtex4-2}

\usepackage[utf8]{inputenc}  
\usepackage{graphicx}
\usepackage{bm}
\usepackage{braket}
\usepackage{xcolor} 
\usepackage{subfigure}     
\usepackage{diagbox}
\usepackage{amsthm}
\usepackage{hyperref}

\definecolor{dkgreen}{rgb}{0,0.6,0}

\definecolor{purple}{rgb}{0.5,0,0.5}

\begin{document}
	
	\title{Quasinormal modes and greybody factor of charged black hole in non-commutative geometry}
	
	\author{Shi-Jie Ma}
	\email[Email: ]{220220939731@lzu.edu.cn} 
	\author{Rui-Bo Wang}
	\email[Email: ]{wangrb2021@lzu.edu.cn}
	\affiliation{Lanzhou Center for Theoretical Physics, Key Laboratory of Theoretical Physics of Gansu Province, Lanzhou University, Lanzhou, Gansu 730000, China}
	\author{Tian-Chi Ma}
	\email[Email: ]{tianchima@buaa.edu.cn}
	\affiliation{Center for Gravitational Physics, Department of Space Science, Beihang University, Beijing 100191, China}
	\author{He-Xu Zhang}
	\email[Email: ]{hxzhang18@163.com}
	\affiliation{Center for Theoretical Physics and College of Physics, Jilin University, Changchun, 130012, China}
	\author{Jian-Bo Deng}
	\email[Email: ]{dengjb@lzu.edu.cn}
	\affiliation{Lanzhou Center for Theoretical Physics, Key Laboratory of Theoretical Physics of Gansu Province, Lanzhou University, Lanzhou, Gansu 730000, China}
	\author{Xian-Ru Hu}
	\email[Email(corresponding author): ]{huxianru@lzu.edu.cn}
	\affiliation{Lanzhou Center for Theoretical Physics, Key Laboratory of Theoretical Physics of Gansu Province, Lanzhou University, Lanzhou, Gansu 730000, China}
	
	\date{\today}
	
	\begin{abstract}
		In this article, the quasinormal modes and greybody factor of charged black hole in non-commutative geometry are studied. Under the assumption of a uniformly distributed charge within the matter, we obtain the metric for a charged black hole in non-commutative geometry. We calculated the wave function and obtained the effective potential of three different perturbed fields with spin. Then we applied $6^{\rm{th}}$ order WKB method to analyze the quasinormal modes of the black hole and derived quasinormal frequencies. Futhermore, we discussed the greybody factor in different perturbed fields under this spacetime. We found that the peak value of effective potential will increase with the increase of the charge of the black holes and the non-commutative parameter in Lorentzian distribution. And the non-commutative parameter in Gaussian distribution has almost no influence. Moreover, the variation trends of the real part and the absolute value of the imaginary part of QNFs, and the greybody factor are exactly same as the variation of the peak value of effective potential.
	\end{abstract}
	
	\maketitle
	\section{Introduction}\label{sec1}
	Non-commutative geometry is a branch of mathematics concerned with a geometric approach to non-commutative algebras, and with the construction of spaces that are locally presented by non-commutative algebras of functions, possibly in some generalized sense. Physicists explore non-commutative spacetime to address the lack of precise boundaries in distance measurements in the formalism of describing spacetime geometry in the domain of general relativity, treating the Planck length as a fundamental constraint~\cite{ncgs_m1,ncgs_m2,ncgs_m3,ncgym1,ncgym2,ncgym3}. The Seiberg-Witten map serves as a common tool for introducing non-commutativity into gravitational theories by gauging an appropriate group~\cite{ncggra}, and this framework has led to significant advancements in understanding black holes. 
	
	Due to non exchange effects, the distribution of matter should be rewritten from a point-like particle to a continous density~\cite{nonbh}. In this process, the normal Einstein equation is preserved. The sperically distribution of matter in three-dimensional space is revised as a Gaussian distribution~\cite{nong} or a Lorentzian distribution~\cite{nonl}. Subsequently, black hole in non-commutative geometry has been extensively studied, including black hole shadow~\cite{bls1,bls2,bls3}, gravitational lensing~\cite{gl1,gl2}, black hole thermodynamics, the Hawking temperature and tunneling effects~\cite{bht1,bht2,ht1,ht2,ht3}, topological features in Gauss-Bonnet gravity~\cite{EGB}, accretion of matter~\cite{aom} and quasinormal modes~\cite{nonqnm1,nonqnm2}
	
	Quasinormal mode (QNM) is a kind of gravitational wave produced by black hole mergers during the ringdown phase~\cite{ringdown}, with a complex characteristic frequency called "quasinormal frequencies" (QNFs). The real parts of the QNFs are the oscillation frequencies of the perturbation, and the imaginary parts are related to the decay time. The QNMs of the gravitational field can only depend on the parameter of the black hole, so the study of QNMs can help us understand black holes more precisely~\cite{infer1,infer2,infer3,infer4}. This is why QNMs have attracted more interest of researchers in recent years. There are many method used to study the QNMs of black holes. The most commonly used method is the Wentzel-Kramers-Brillouin (WKB) approximative method~\cite{CF,DI,WKB1,WKB2,WKB3,WKB4,QNMtest,4DEGB}. 
	
	In addition, greybody factor is also a significant concept for black hole spacetime with perturbed field. Greybody factor is used to describe the transmission probability of an outgoing wave reaching to infinity or an incoming wave to be absorbed by black hole~\cite{gbf1,gbf2,gbf3}. The greybody factor is so important because it could effectively help to analyze the information behavior near horizon regions of a black hole~\cite{gbf4}. And it could be used to estimate the energy emission of Hawking radiation~\cite{gbf5}. Besides, it should be emphasized that it was suggested that greybody factor played a important role in the research on ringdown signal after an extreme mass ratio merger~\cite{gbf6}.
	
	At present, there are many non-commutative black holes' properties investigated by the researchers. But for the charged black holes under the background of non-commutative geometry, there remains to be studied in detail. So in this paper, we construct a toy model of charged black hole in non-commutative geometry, studying its QNM and greybody factor. This article is organized as follows. In Sect.~\ref{sec2}, we derive the metric of charged black hole in non-commutative geometry. In Sect.~\ref{sec3}, we discuss the Klein-Gordon equation of QNMs and the effective potential. In Sect.~\ref{sec4}, the $6^{\rm{th}}$ order WKB method is applied to study the QNFs and the influences of the electric charge and the non-commutative parameter. In Sect.~\ref{sec5}, we calculate the greybody factor in detail. The conclusion will be given in Sect.~\ref{sec6}. Our work will provide a meaningful reference for the study of gravitational aspect in non-commutative geometry. In this article, we use the natural units $G_N=c=4\pi\varepsilon_{0}=\hbar=1$.
	
	\section{charged black hole in non-commutative geometry}\label{sec2}
	The Einstein equation is 
	\begin{equation}
		R^{\nu}_{\mu}-\frac{1}{2}\delta^{\nu}_{\mu}R=8\pi T^{\nu}_{\mu},
	\end{equation}
	where $R^{\nu}_{\mu}$ is Ricci tensor, $\delta^{\nu}_{\mu}$ is Kronecker symbol, $R$ is Ricci scalar and $\delta^{\nu}_{\mu}$ is energy-momentum tensor. For a static spherically symmetric black hole, its metric could be written as
	\begin{equation}\label{eq1}
		ds^2=-f\left(r\right)dt^2+\frac{1}{f\left(r\right)}dr^2+r^2 \left(d\theta^2+\sin^2\theta d\varphi^2\right).
	\end{equation}
	By substituting the metric into the Einstein equation, one could obtain
	\begin{equation}
		f\left(r\right)=1+\frac{\int 8\pi r^{2} T_{0}^{0}dr}{r},
	\end{equation}
	and
	\begin{equation}
		T_{0}^{0}=\left\{
		\begin{array}{clc}
			-\rho,&\qquad &\text{matter}, \\
			
			-\frac{1}{2}\left( E^2+B^2\right),
			&\qquad& \text{electromagnetic\ field},
		\end{array}
		\right.
	\end{equation}
	where $\rho$ is the distribution function of matter. $E$ and $B$ are the electric field strength and the magnetic induction strength, respectively.
	
	In non-commutative geometry, the distribution of matter in space can be considered as a Gaussian distribution or a Lorentzian distribution~\cite{nong,nonl}.
	\begin{equation}
		\rho_{M}=\left\{
		\begin{array}{rc}
			\frac{M}{\left(4\pi \Theta\right)^\frac{3}{2}}e^{-\frac{r^2}{4\Theta}}, \qquad & \text{Gaussian\ distribution}, \\
			
			\frac{M\sqrt{\Theta}}{\pi^\frac{3}{2}\left(r^2+\pi\Theta\right)^2}, \qquad & \text{Lorentzian\ distribution},
		\end{array}
		\right.
	\end{equation}
	where $M$ is the total mass and $\Theta$ is the non-commutative parameter with dimension of $\left[\rm{L}\right]^2$. Assuming that the charge is uniformly distributed on the matter, one could obtain the distribution of charge
	\begin{equation}
		\rho_{Q}=\left\{
		\begin{array}{rc}
			\frac{Q}{\left(4\pi \Theta\right)^\frac{3}{2}}e^{-\frac{r^2}{4\Theta}}, \qquad & \text{Gaussian\ distribution}, \\
			
			\frac{Q\sqrt{\Theta}}{\pi^\frac{3}{2}\left(r^2+\pi\Theta\right)^2}, \qquad & \text{Lorentzian\ distribution},
		\end{array}
		\right.
	\end{equation}
	with $Q$ is the total charge.
	
	According to Maxwell's equations, one can obtain
	\begin{equation}
		B\left(r\right)=0,\qquad E\left(r\right)=\frac{\int_{0}^{r}4\pi \tilde{r}^2\rho_{Q}d\tilde{r}}{r^2}.	\end{equation}
	
	One can substitute this into the Einstein equation, and further make a transform $r/M\rightarrow r$, $\Theta/M\rightarrow \theta$, $Q/M\rightarrow q$ to obtain the metric of charged black hole in non-commutative geometry
	
	\begin{equation}
		f_{g}\left(r\right)=
		1-\frac{2}{r}\mathrm{Erf}\left(\frac{r}{2\sqrt{\theta}}\right)
		+\frac{q^{2}}{r^{2}}\mathrm{Erf}\left(\frac{r}{2\sqrt{\theta}}\right)^{2}
		+\frac{2e^{-\frac{r^2}{4\theta}}}{\sqrt{\pi\theta}}
		+\frac{q^2\left(1-\mathrm{Erf}\left(\frac{r}{\sqrt{2\theta}}\right)\right)}{r\sqrt{2\pi\theta}},
	\end{equation}
	\begin{equation}
		f_{l}\left(r\right)=
		1+\frac{4\sqrt{\theta}}{\sqrt{\pi}\left(\theta\pi+r^2\right)}
		-\frac{2q^2}{\pi^2\left(\theta\pi+r^2\right)}
		+\left(\frac{2q~\mathrm{arccot}\left(\frac{\sqrt{\pi\theta}}{r}\right)}{\pi r}\right)^{2}
		+\frac{2q^2\arctan\left(\frac{\sqrt{\pi\theta}}{r}\right)}{\sqrt{\pi^5\theta}r}
		-\frac{4\arctan\left(\frac{r}{\sqrt{\pi\theta}}\right)}{\pi r},
	\end{equation}
	
	where $\mathrm{Erf}\left(x\right)=\frac{2}{\sqrt{\pi}}\int_{0}^{x}e^{-t^{2}}dt$ is the error function.
	
	\begin{center}
		\begin{figure}[htbp] 
			\includegraphics[width=1\textwidth]{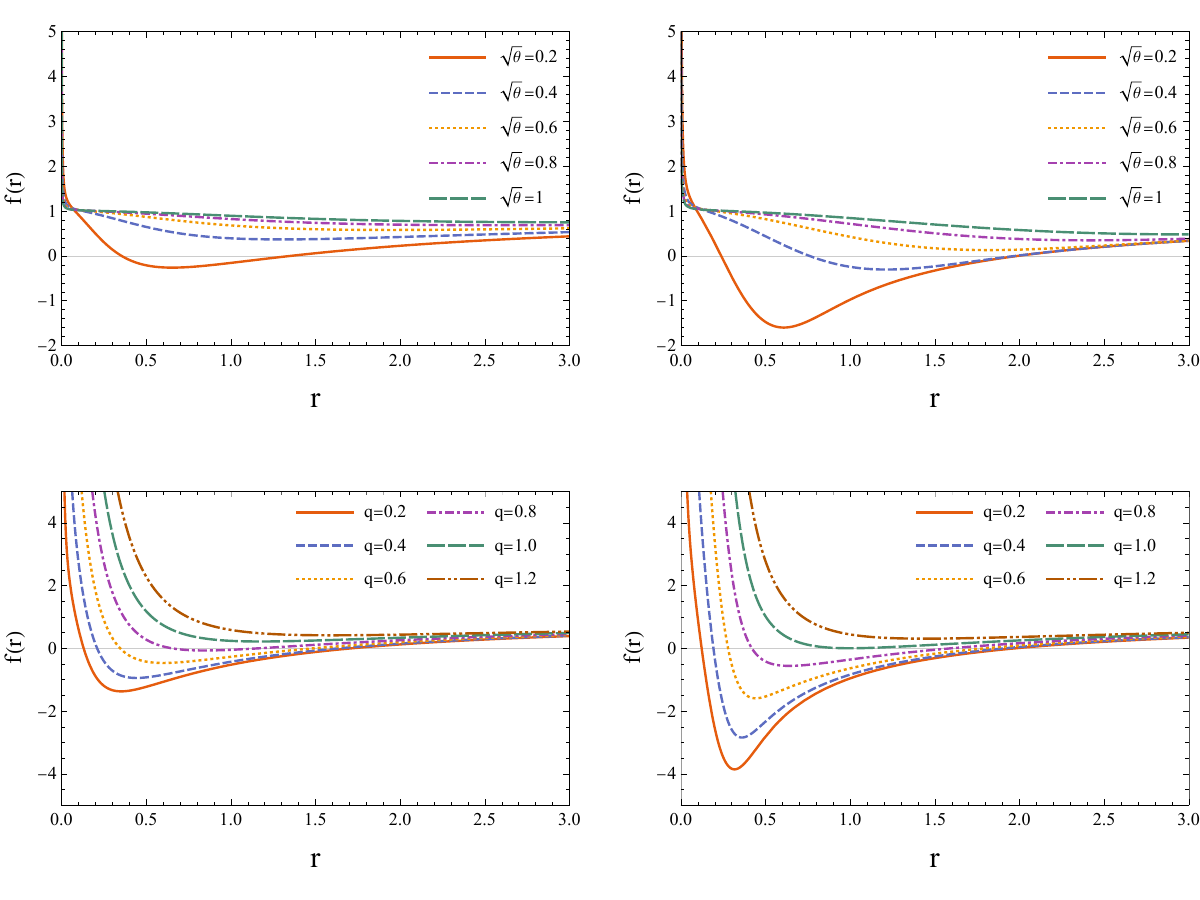}
			\caption{The black hole's metric $f\left(r\right)$ of  Lorentzian distribution (left) and Gaussian distribution (right) for different $q$ and $\sqrt{\theta}$.}\label{figPf}
		\end{figure}
	\end{center}
	
	From Fig.~\ref{figPf}, it can be observed that when the parameter $\sqrt{\theta}$ or $q$ is sufficiently large, no solution exists for $f\left(r_{h}\right)=0$ in either distribution, implying the absence of a black hole solution. Moreover, it is evident that for a fixed $q$ ($\sqrt{\theta}$), there exists a critical value of $\sqrt{\theta}$ ($q$) at which the $f(r)$ curve is exactly tangent to the $r$-axis. However, these critical values cannot be obtained analytically and provide limited insight for our subsequent research. Therefore, we do not discuss them further. Further, from the figure, an increase of $\sqrt{\theta}$ leads to a more dispersed representation of matter, resulting in insufficient gravitational force. An increase of $q$ indicates an increase in charges within the Gaussian plane, leading to a stronger repulsive force.
	
	\section{perturbed field and effective potential}\label{sec3}
	
	With the background spacetime of a static spherically symmetric black hole, a massless scalar field $\Phi$ obeys the Klein-Gorden equation
	\begin{equation}\label{eq2}
		\Box\Phi\equiv\frac{1}{\sqrt{-g}}\partial_{\mu}\left(\sqrt{-g}g^{\mu\nu}\partial_{\nu}\Phi\right)=0,
	\end{equation}
	where $g^{\mu\nu}$ is inverse metric, $g$ is the determinant of metric $g_{\mu\nu}$, and $\Phi$ is the wave function incoming from infinity towards the black hole.
	
	As is done in quantum mechanics, the Klein-Gordon equation could be solved by separating the variables
	\begin{equation}\label{eq3}
		\Phi\left(t,r,\vartheta,\phi\right)=\frac{e^{-i\omega t}}{r}\Psi\left(r\right)Y_{\ell}^{m}\left(\vartheta,\phi\right),
	\end{equation}
	where $\omega$ is the frequency of $\Phi$, $\Psi\left(r\right)$ is radial wave function. $\ell$ is azimuthal quantum number, and takes only nonnegative integers. $m$ is magnetic quantum number, and in this article we always choose it as zero. $Y_{\ell}^{m}$ is spherical harmonic function, and it fulfills the relations
	
	\begin{equation}\label{eq4}
		\left[\frac{1}{\sin\vartheta}\partial_{\vartheta}\left(\sin\vartheta\partial_{\vartheta}\right)+ \frac{1}{\sin^2\vartheta}\partial_{\phi}^2\right]Y_{\ell}^{m}\left(\vartheta,\phi\right)=-\ell(\ell+1)Y_{\ell}^{m}\left(\vartheta,\phi\right).
	\end{equation}
	
	Inserting Eq.~\eqref{eq3} into Eq.~\eqref{eq2}, the Regge-Wheeler wave equation~\cite{RW} is obtained
	\begin{equation}\label{eq5}
		\left[\frac{\partial^{2}}{\partial t^{2}}-\frac{\partial^2}{\partial r_{\ast}^2}+V_{eff}
		\left(r\right)\right]\Psi\left(r\right)e^{-i\omega t}=0,
	\end{equation}
	which could also be written as
	\begin{equation}\label{eq6}
		\frac{d^2}{dr_{\ast}^{2}}\Psi+\left(\omega^2-V_{eff}\right)\Psi=0,
	\end{equation}
	with ”tortoise” radial coordinate $r_{\ast}=\int dr/f\left(r\right)$ and the effective potential $V_{eff}$.  For a massless scalar field, $V_{eff}$ reads
	\begin{equation}\label{eq7}
		V_{eff}\left(r\right)=f\left(r\right)\left[\frac{\ell\left(\ell+1\right)}{r^2}+\frac{1}{r}\frac{df\left(r\right)}{dr}\right].
	\end{equation}
	A generalized form of the effective potential for higher spin (boson) fields is given by~\cite{veff1,veff2}
	
	\begin{equation}\label{eq8}
		V_{eff}\left(r\right)=f\left(r\right)\left[\frac{\ell\left(\ell+1\right)}{r^2}+\left(s-s^2\right)\frac{1-f\left(r\right)}{r^2}+\left(1-s\right)\frac{1}{r}\frac{df\left(r\right)}{dr}\right],
	\end{equation}
	
	where $s\leq l$ is the spin of the perturbative field
	\begin{equation}\label{eq9}
		s=\left\{
		\begin{array}{rc}
			0, \qquad & \text{scalar\ perturbation}, \\
			1, \qquad & \text{electromagnetic\ perturbation}, \\
			2, \qquad & \text{gravitational\ perturbation}.
		\end{array}
		\right.
	\end{equation}
	
	\begin{center}
		\begin{figure}[htbp]  
			\centering
			\includegraphics[width=1\textwidth]{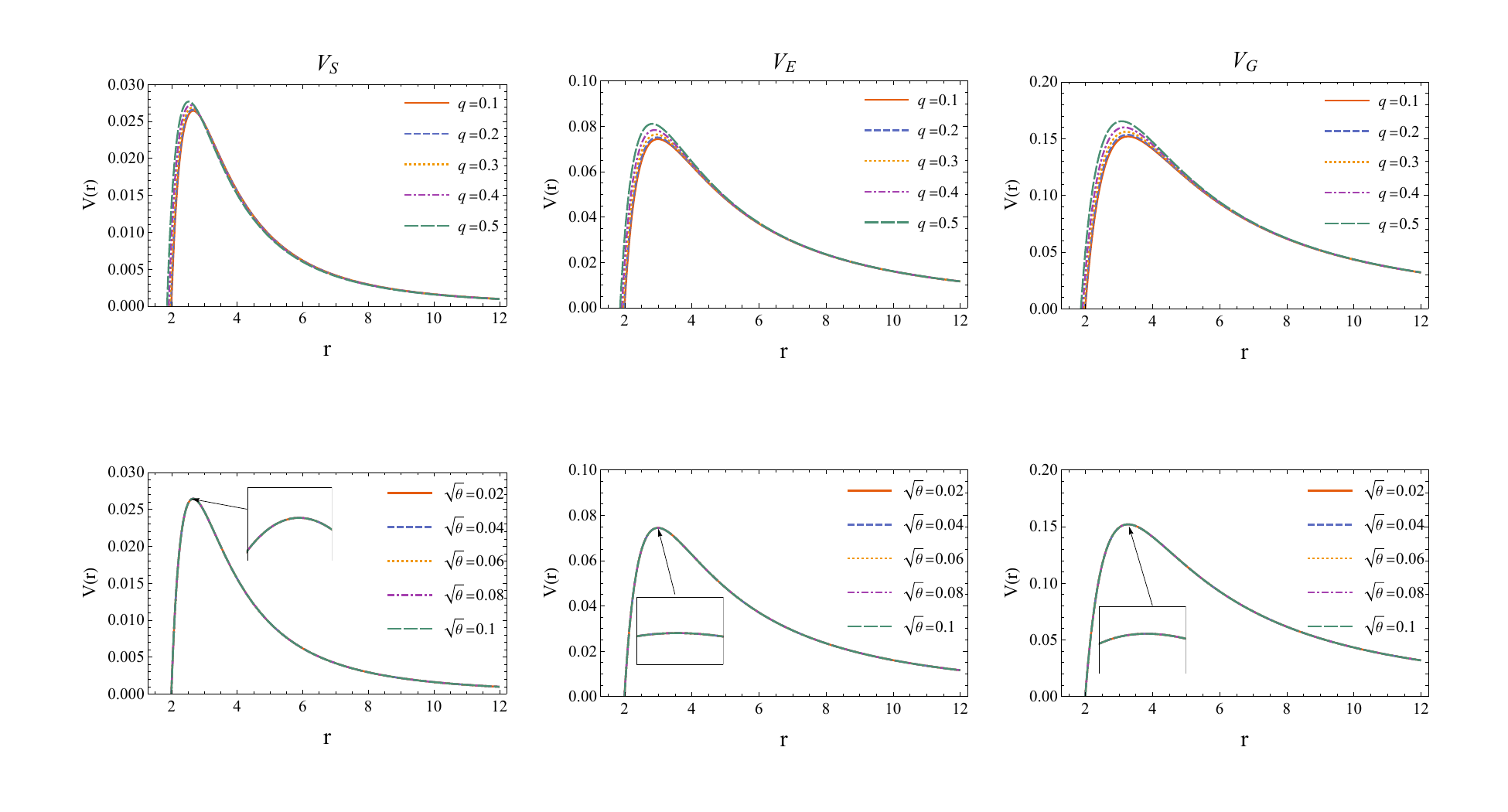}
			\caption{$V_{eff}$ of Gaussian distribution for different perturbed fields $V_S$ (scalar field) at $\ell=0$, $V_E$ (electromagnetic field) at $\ell=1$, and $V_G$ (gravitational field) at $\ell=2$, respectively. The first row is $V_{eff}$ for changing $q$ at $\sqrt{\theta}=0.1$, the second row is $V_{eff}$ for changing $\sqrt{\theta}$ at $q=0.1$.}\label{figVg}
		\end{figure}
	\end{center}
	
	The plots for $V_{eff}$ of Gaussian distribution for three perturbed fields are shown in Fig.~\ref{figVg}. From these figures one can see that the effect of the parameter $q$ on the effective potentials for three perturbed fields are the same, that is the heights of the potentials increase with $q$. However, the impact of $\sqrt{\theta}$ is almost negligible.
	
	\begin{center}
		\begin{figure}[htbp]  
			\centering
			\includegraphics[width=1\textwidth]{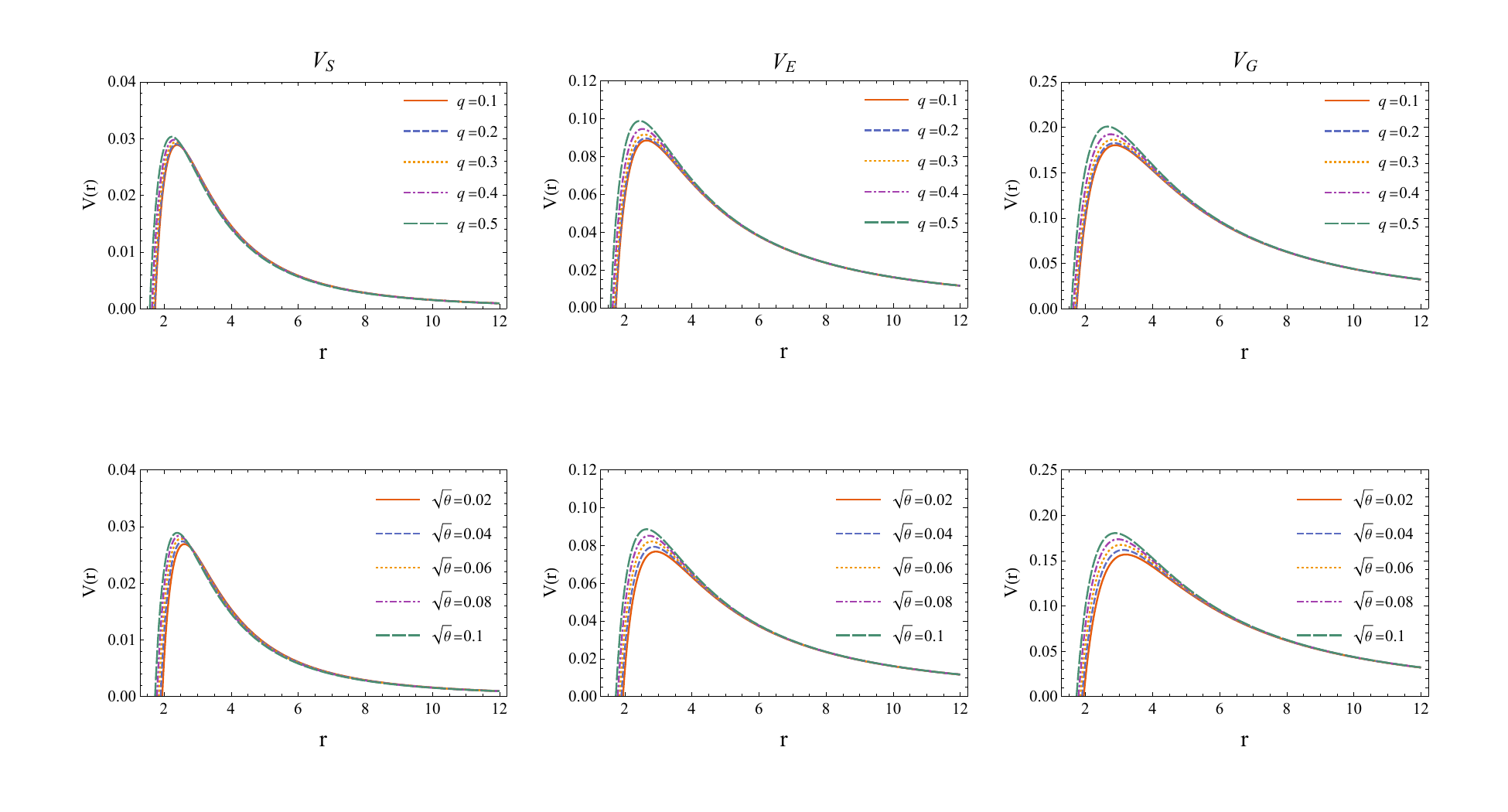}
			\caption{$V_{eff}$ of Lorentzian distribution for different perturbed fields $V_S$ at $\ell=0$, $V_E$ at $\ell=1$, and $V_G$ at $\ell=2$, respectively. The first row is $V_{eff}$ for changing $q$ at $\sqrt{\theta}=0.1$, the second row is $V_{eff}$ for changing $\sqrt{\theta}$ at $q=0.1$.}\label{figVl}
		\end{figure}
	\end{center}

	Fig.~\ref{figVl} shows the plots for $V_{eff}$ of Lorentzian distribution for three perturbed fields. From these figures one can see that, the effect of the parameters $q$ and $\sqrt{\theta}$ on the effective potentials for three perturbed fields are the same, that is the heights of the potentials increase with $q$ and $\sqrt{\theta}$.
	
	\section{Quasinormal modes and WKB method}\label{sec4}
	
	The region to be studied should be outside the event horizon, that means $r>r_{+}$ or $r_{\ast}\in\left(-\infty,+\infty\right)$. The boundary condition for this problem is
	\begin{equation}\label{eq11}
		\Psi\propto \left\{\begin{array}{lr}
			e^{-i\omega r_{\ast}},\qquad\qquad\qquad & r_{\ast}\rightarrow-\infty,\\
			e^{i\omega r_{\ast}},\qquad\qquad\qquad & r_{\ast}\rightarrow+\infty,
		\end{array}\right.
	\end{equation}
	which is only an ingoing wave at the horizon and only an outgoing wave at infinity. 
	
	The WKB method is based on matching of the asymptotic solutions, with the Taylor expansion around the peak of $V_{eff}$. By this way, it is possible to relate the ingoing and outgoing amplitudes through the linear transformation. Then, one could obtain $k^{\rm{th}}$ order WKB formula for solving the QNFs
	\begin{equation}\label{eq12}
		\frac{i\left(\omega^2-V_{0}\right)}{\sqrt{-2V_{2}}}+\sum_{j=2}^{k}\Lambda_{j}=n+\frac{1}{2},
	\end{equation}
	where $n$ is principal quantum number, $V_{j}$ are set as the $j^{\rm{th}}$ derivative with respect to the $r_{\ast}$ at the peak value of the $V_{eff}$ ($j\geq0$, defining the "$0^{\rm}$ derivative" as the original, non-differentiated function), and $\Lambda_{j}$ are the correction terms depending on $V_{0}$ to $V_{2j}$~\cite{WKB1,WKB2,WKB3,WKB4}. One can obtain the change trend of QNFs by simply analyzing $1^{\rm{st}}$ order WKB equation. When $V_0$ or $n$ increases, the real part and imaginary part of $\omega^2$ will increase. The real parts of QNFs $\omega_{r}$ are positive and the imaginary parts of QNFs $\omega_{i}$ are negative~\cite{QNMtest}, so it can be concluded that $\left|\omega_{r}\right|$ and $\left|\omega_{i}\right|$ both increase as $V_{eff}$ or $n$ increases, and the increment of $\left|\omega_{r}\right|$ is greater than that of $\left|\omega_{i}\right|$. 
	
\begin{center}
		\begin{figure}[htbp]
			\includegraphics[width=\textwidth]{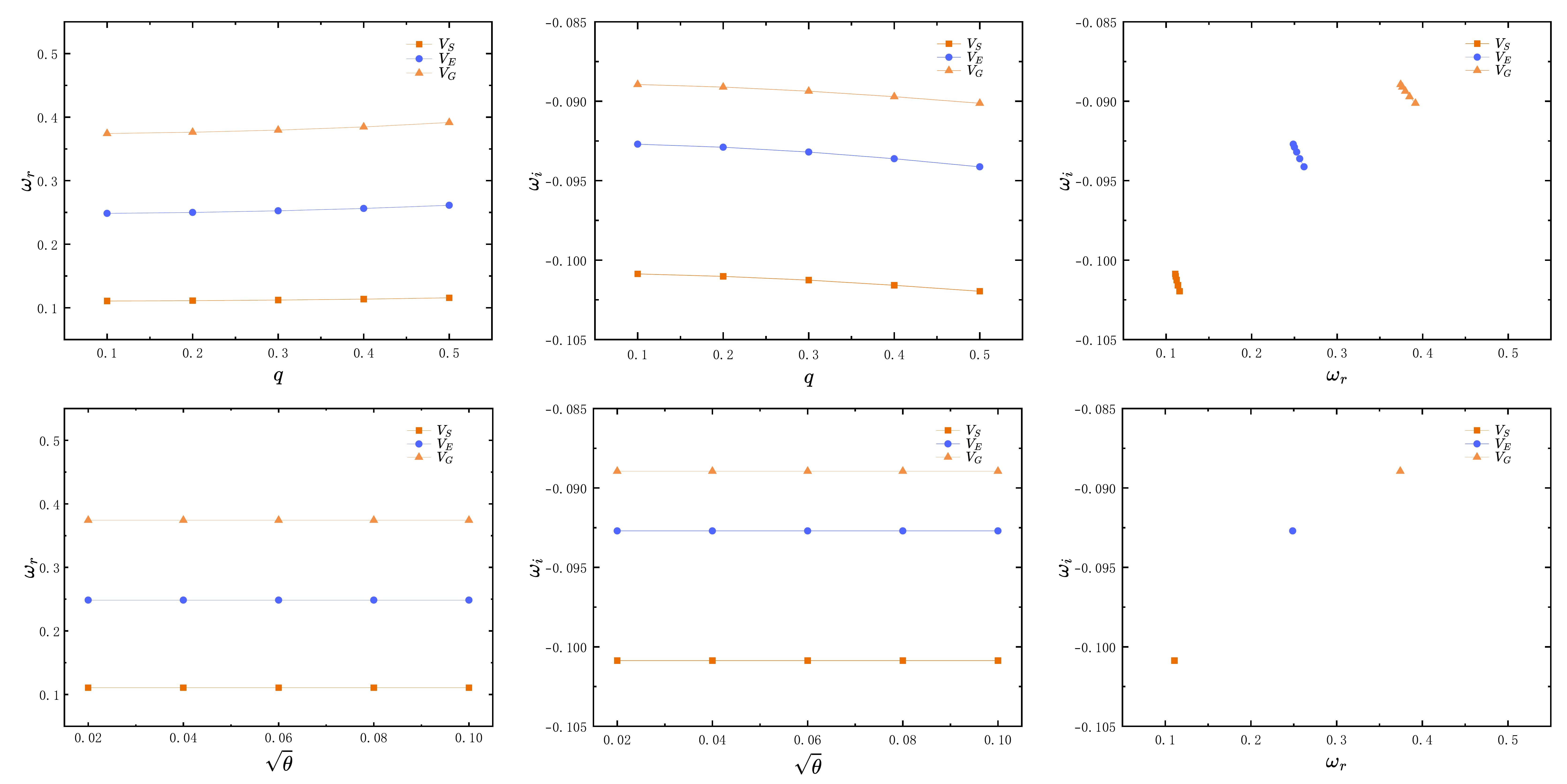}
			\caption{The QNFs of Gaussian distribution for different perturbed fields $V_S$ at $\ell=0$, $V_E$ at $\ell=1$, and $V_G$  at $\ell=2$, respectively. The first row plots are $V_{eff}$ for changing $q$ at $\sqrt{\theta}=0.1$, the second row plots are $V_{eff}$ for changing $\sqrt{\theta}$ at $q=0.1$. Set $n=0$.}\label{figog}
		\end{figure}
	\end{center}

\begin{center}
		\begin{figure}[htbp]
			\includegraphics[width=\textwidth]{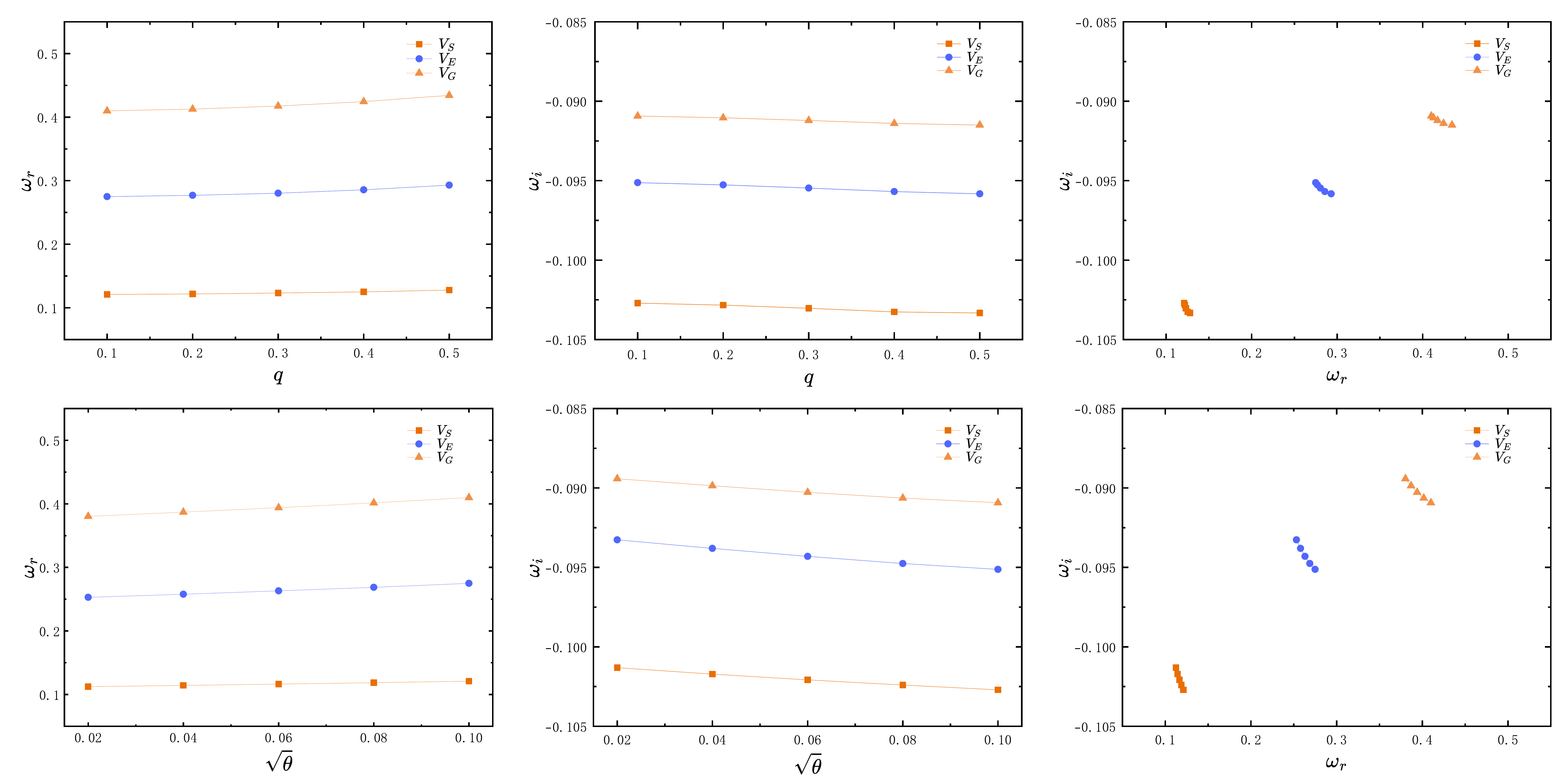}
			\caption{The QNFs of Lorentzian distribution for different perturbed fields $V_S$ at $\ell=0$, $V_E$ at $\ell=1$, and $V_G$  at $\ell=2$, respectively. The first row plots are $V_{eff}$ for changing $q$ at $\sqrt{\theta}=0.1$, the second row plots are $V_{eff}$ for changing $\sqrt{\theta}$ at $q=0.1$. Set $n=0$.}\label{figol}
		\end{figure}
	\end{center}

	The WKB method is a kind of approximation methods, and there is a small error between the obtained result and the exact solution. With the increase of the approximation order, the calculation will be more accurate. To analyze the impact of changes in $q$ and $\sqrt{\theta}$ on QNFs, the commonly used $6^{\rm{th}}$ order WKB method is accurate enough. The QNFs of Gaussian distribution and Lorentzian distribution with different values of $q$ and $\sqrt{\theta}$ are plotted in Fig.~\ref{figog} and Fig.~\ref{figol}, respectively.
	
	Fig.~\ref{figog} shows the QNFs of Gaussian distribution. As $q$ increases, the absolute values of $\left|\omega_{r}\right|$ and $\left|\omega_{i}\right|$ of QNfs both increase. The increment of $\left|\omega_{r}\right|$ is greater than that of $\left|\omega_{i}\right|$. Because the $V_{eff}$ effect of $\sqrt{\theta}$ on Gaussian distribution is extremely small, its impact on QNFs is also almost indistinguishable, appearing as only one point in the figure.
	
	The QNFs of Lorentzian distribution are showed in Fig.~\ref{figol}. As $q$ and $\sqrt{\theta}$ increase, the absolute values of $\left|\omega_{r}\right|$ and $\left|\omega_{i}\right|$ both increase. The increment of $\left|\omega_{r}\right|$ is more apparent than that of $\left|\omega_{i}\right|$.

	\section{The greybody factor}\label{sec5}
	
	Just as tunneling effect in quantum mechanics, the reflection and transmission of waves through black hole spacetime should also be considered. There are both ingoing wave and outgoing wave at infinity, but only ingoing wave at horizon. Thus, the boundary condition should be written as
	\begin{equation}\label{eq16}
		\Psi=\left\{\begin{array}{lr}
			Te^{-i\omega r_{\ast}},\qquad\qquad\qquad & r_{\ast}\rightarrow-\infty,\\
			e^{-i\omega r_{\ast}}+R e^{i\omega r_{\ast}},\qquad\qquad\qquad & r_{\ast}\rightarrow+\infty,
		\end{array}\right.
	\end{equation}
	where $R$ and $T$ represent the reflection coefficient and the transmission coefficient, respectively. The greybody factor is defined as the probability of an outgoing wave reaching to infinity or an incoming wave absorbed by the black hole~\cite{gbf1,gbf2,gbf3}, so $\left|T\right|^2$ is called the greybody factor. $R$ and $T$ should satisfy the following relation
	\begin{equation}\label{eq17}
		\left|R\right|^2+\left|T\right|^2=1.
	\end{equation}
	\begin{center}
		\begin{figure}[htpb]
			\centering
			\includegraphics[width=0.46\textwidth]{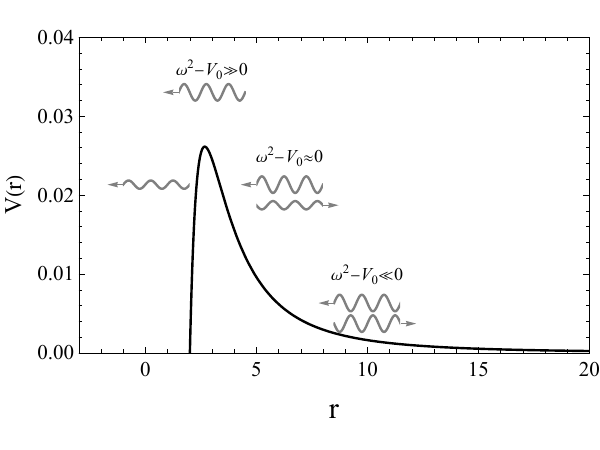}
			\caption{Classification diagram of gravitational wave transmission scenarios.}\label{figov}
		\end{figure}
	\end{center}
	\begin{table}[htpb]
		\caption{The relationship between the result of $\omega^2-V_{0}$ and whether the reflection or transmission effect is significant. \checkmark means the transmission effect is obvious, $\times$~means it is not obvious.}\label{tab4}
		\centering
		\begin{tabular}{ccc}
			\hline
			&reflection&transmission\\
			\hline
			$\omega^2-V_{0}\ll 0$&\checkmark&$\times$\\
			$\omega^2-V_{0}\approx 0$&\checkmark&\checkmark\\
			$\omega^2-V_{0}\gg 0$&$\times$&\checkmark\\
			\hline
		\end{tabular}
	\end{table}
	Fig.~\ref{figov} and Table~\ref{tab4} show that when $\omega^2-V_{0}\approx 0$, the transmission effect and reflection effect are more obvious. This is obviously a condition that satisfies the solution of the WKB method. Using the WKB method~\cite{gbf_WKB}, the reflection and transmission coefficients could be obtained
	\begin{equation}\label{eq18}
		\left|R\right|^2=\frac{1}{1+e^{-i2\pi K}},
	\end{equation}
	\begin{equation}\label{eq19}
		\left|T\right|^2=\frac{1}{1+e^{+i2\pi K}},
	\end{equation}
	\begin{equation}\label{eq20}
		K=\frac{i\left(\omega^2-V_{0}\right)}{\sqrt{-2V_{2}}}+\sum_{j=2}^{k}\Lambda_{j},
	\end{equation}
	where $K$ is a parameter obtained by the WKB formula and $\Lambda_{j}$ are all imaginary numbers.
	
	\begin{center}
		\begin{figure}[htpb]
			\centering
			\includegraphics[width=\textwidth]{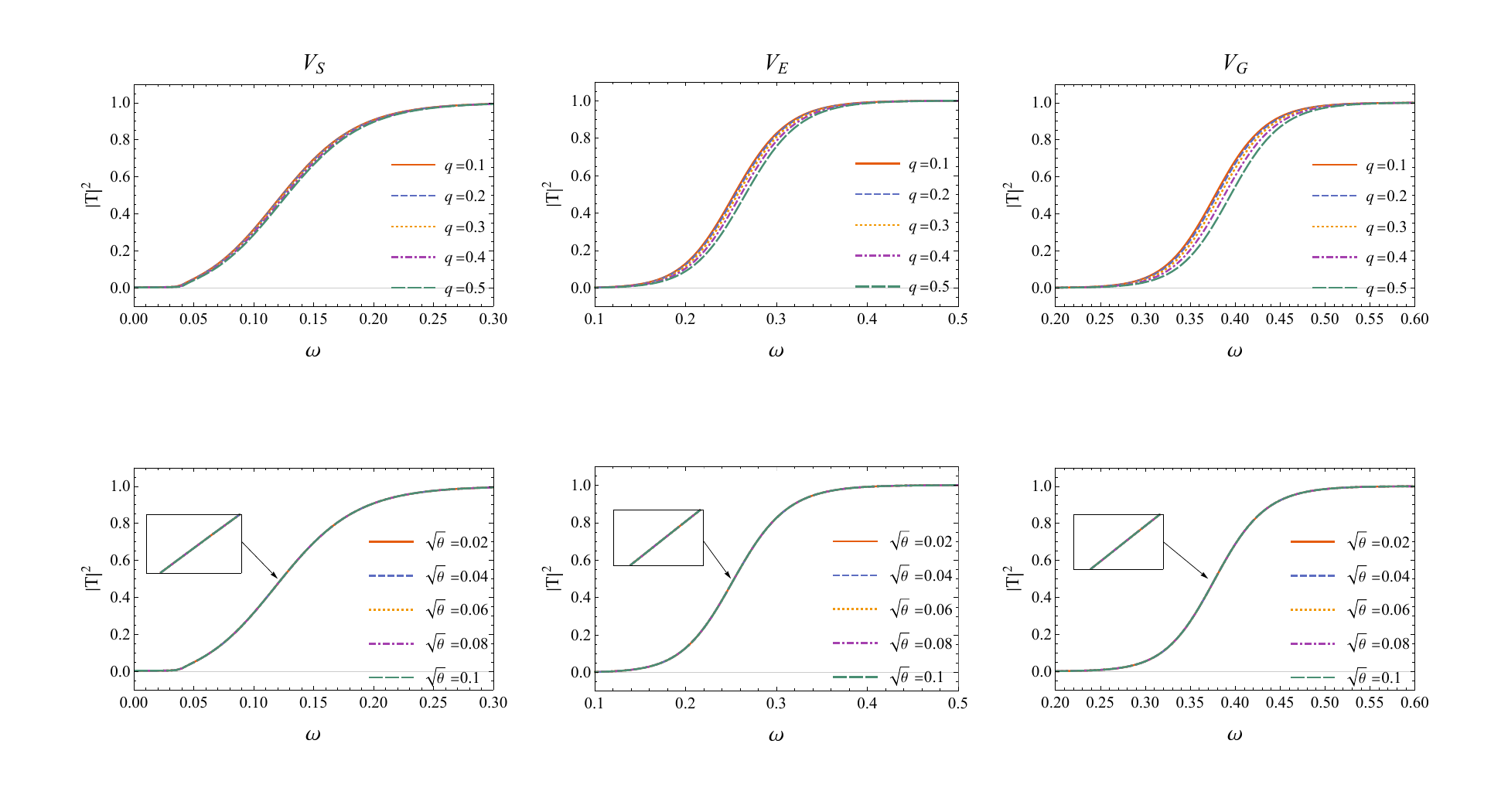}
			\caption{The $\left|T\right|^2$ of Gaussian distribution for different perturbed fields $V_S$ at $\ell=0$, $V_E$ at $\ell=1$, and $V_G$  at $\ell=2$, respectively. The first row plots are $V_{eff}$ for changing $q$ at $\sqrt{\theta}=0.1$, the second row plots are $V_{eff}$ for changing $\sqrt{\theta}$ at $q=0.1$.}\label{figgtg}
		\end{figure}
		\begin{figure}[htpb]
			\centering
			\includegraphics[width=\textwidth]{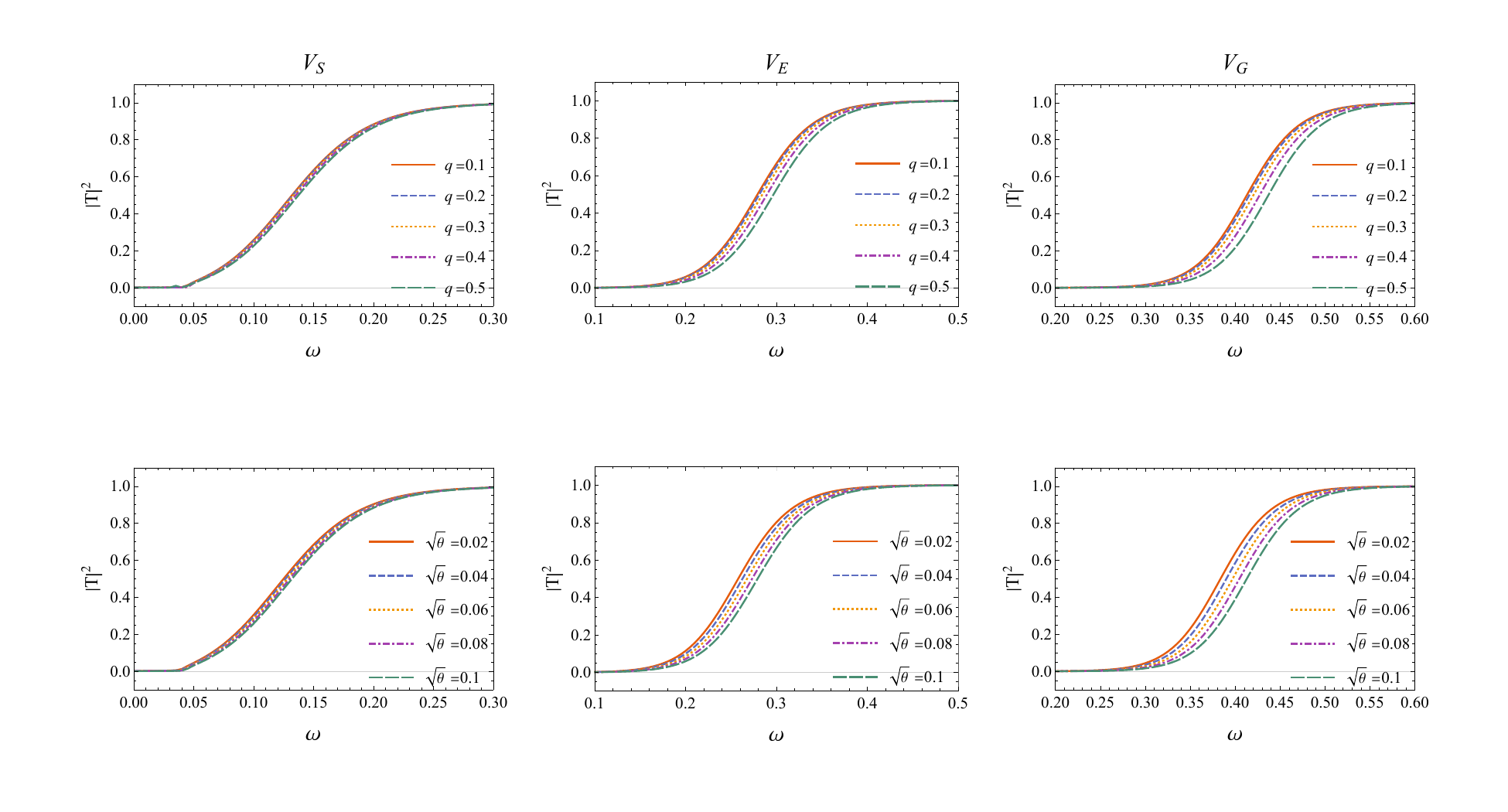}
			\caption{The $\left|T\right|^2$ of Lorentzian distribution for different perturbed fields $V_S$ at $\ell=0$, $V_E$ at $\ell=1$, and $V_G$  at $\ell=2$, respectively. The first row plots are $V_{eff}$ for changing $q$ at $\sqrt{\theta}=0.1$, the second row plots are $V_{eff}$ for changing $\sqrt{\theta}$ at $q=0.1$.}\label{figgtl}
		\end{figure}
	\end{center}

	Fig.~\ref{figgtg} shows that $\omega$ decreases as $q$ increases at the same $\left|T\right|^2$, but the impact of $\sqrt{\theta}$ is almost negligible. From Fig.~\ref{figgtl}, one can see that $\omega$ decreases as $q$ or $\sqrt{\theta}$ increases at the same $\left|T\right|^2$. This is consistent with the discussion of effective potential, where the height of the potential barrier decreases and the perturbation could pass through the barrier more easily as $q$ or $\sqrt{\theta}$ decreases.
	
	\section{Conclusion and outlook}\label{sec6}
	
	Assuming that the charge is uniformly distributed in the material, by solving the field equation, we obtained the metric of charged black holes in non-commutative geometry with Gaussian distribution and Lorentzian distribution of matter, firstly. Then we studied the perturbation equations of three kinds of perturbation fields with different spins in the background of the charged black hole in non-commutative geometry. The effective potential is calculated and plotted. According to this, it can be concluded that for Gaussian distribution, as the charge $q$ increases, the peak of the effective potential increases, and the non exchange parameter $\sqrt{\theta}$ has almost no effect; for the Lorentzian distribution, the peak of the effective potential increases with the increase of $q$ or $\sqrt{\theta} $. After applying the boundary condition, we calculated QNFs using $6^{\rm{th}}$ order WKB method. The results shown that since QNFs are solved near the peak of the effective potential, the results of QNFs are affected by changes in the effective potential. For Gaussian distribution, the real parts and the absolut value of the imaginary parts of the QNFs increase as $q$ increases, but the effect of $\sqrt{\theta}$ variation on QNFs is minimal, consistent with the effective potential. For Lorentzian distribution, the real parts and the absolut value of the imaginary parts of the QNFs increase as $q$ or $\sqrt{\theta}$ increases.  Finally, we also discussed the greybody factor for these three perturbed fields by used the WKB method. Similarly, influenced by the effective potential, the greybody factor of Gaussian distribution increases with the increase of $q$, but the influence of $\sqrt{\theta}$ can be ignored; The greybody factor of the Lorentzian distribution increases with the increase of $q$ or $\sqrt{\theta}$. It is our sincere hope that this research could proveide a meaningful reference for the studies of theory of gravity under the background of non-commutative geometry.
	
	In this article, we only studied the quasinormal scale and greybody factor of charged black holes in non-commutative geometry. In the future, we will also study the optics and thermodynamics of this kind of black holes. The rotating black holes and black holes with magnetic charges in non-commutative geometry are also interesting subjects.
	
	\section*{Conflicts of interest}
	The authors declare that there are no conflicts of interest regarding the publication of this paper.
	
	\section*{Acknowledgments}
	We want to thank Lei-You, Yu-Cheng Tang and Yu-Hang Feng for thier valuable suggestions.
	
	\section*{Data availability}
	No Data associated in the manuscript.

	\bibliography{paper}
	
\end{document}